# 1/*f* Noise in Thin Films of Topological Insulator Materials

**M. Zahid Hossain[1], Sergey L. Rumyantsev[2,3], Desalegne Teweldebrhan[1], Khan M. F. Shahil[1], Michael Shur[2] and Alexander A. Balandin[1,*]**

[1] Nano-Device Laboratory, Department of Electrical Engineering and Materials Science and Engineering Program, Bourns College of Engineering, University of California – Riverside, Riverside, California 92521 USA
[2] Department of Electrical, Computer and Systems Engineering and Center for Integrated Electronics, Rensselaer Polytechnic Institute, Troy, New York 12180 USA
[3] Ioffe Institute, Russian Academy of Sciences, St. Petersburg, 194021 Russia

* Corresponding author: balandin@ee.ucr.edu, Phone: +7-951-827-2351, URL: http://ndl.ee.ucr.edu/

We report results of investigation of the low-frequency excess noise in device channels made from topological insulators – a new class of materials with a bulk insulating gap and conducting surface states. The thin-film $Bi_2Se_3$ samples were prepared by the "graphene-like" mechanical exfoliation from bulk crystals. The fabricated four-contact devices had linear current – voltage characteristics in the low-bias regime $|V_{SD}|<0.1$ V. The current fluctuations had the noise spectral density $S_I \sim 1/f$ for the frequency $f$ <10 kHz. The noise density $S_I$ followed the quadratic dependence on the drain – source current and changed from about ~$10^{-22}$ to $10^{-18}$ $A^2$/Hz as the current increases from ~$10^{-7}$ to $10^{-5}$ A. The obtained data is important for planning transport experiments with topological insulators. We suggest that achieving the pure topological insulator phase with the current conduction through the "protected" surface states can lead to noise reduction via suppression of certain scattering mechanisms. The latter has important implications for implementing the ultra-low-power and ultra-low-noise electronics.

**1 Introduction** Topological insulators are the materials with a bulk insulating gap and conducting surface states that are topologically protected against scattering by the time-reversal symmetry [1]. This newly discovered class of materials was predicted to reveal many unique properties, e.g. quantum-Hall-like behaviour in the absence of magnetic field. Some of these properties have already been demonstrated, stimulating interest to topological insulators as possible materials for quantum computing and magnetic memory with pure electrical read-write operations.

Bismuth Selenide ($Bi_2Se_3$) and related thin films are becoming the reference topological materials. $Bi_2Se_3$ has a relatively large band gap of ~0.3 eV and a single surface state that has Dirac-cone type dispersion. In addition, $Bi_2Se_3$ is stoichiometric, unlike some other alloy topological insulators, and can be prepared with high purity and low disorder. Although the topological insulator phase was predicted to be robust to disorder, the experimental evidence suggests that crystallinity, high-purity and absence of defects are essential [2]. The naturally occurring defects can lead to the situation when the Fermi level falls in either the conduction or valence bands resulting in a mixed bulk (i.e. volume) and surface conduction.

The importance of the distinction between the volume and surface conduction in topological insulators brings up analogies with such a ubiquitous phenomenon as 1/*f* noise, which could be linked to either volume or surface conduction or both. The low-frequency fluctuations in electrical current, commonly known as "excess" or "flicker" noise, have a well-defined 1/*f* spectral density for the frequency *f* below ~10 – 30 kHz [3]. This type of noise is detected as fluctuations of voltage across a resistor carrying a constant current. It has been observed essentially in all solid state materials including metals and semiconductors and in all solid state devices. The 1/*f* noise limits performance of electronic communication systems and sensors. Despite its practical importance, the origin of 1/*f* noise is still the subject of considerable debates. The question of fundamental importance is whether 1/*f* noise has the bulk (volume) or surface origin [4-5]. For this reason, studies of 1/*f* noise in materials and thin films, which were identified as topological insulators, are of crucial importance. In this letter, we report on the results of the 1/*f* noise measurements in $Bi_2Se_3$ thin films and discuss the implications of our results for topological insulators.

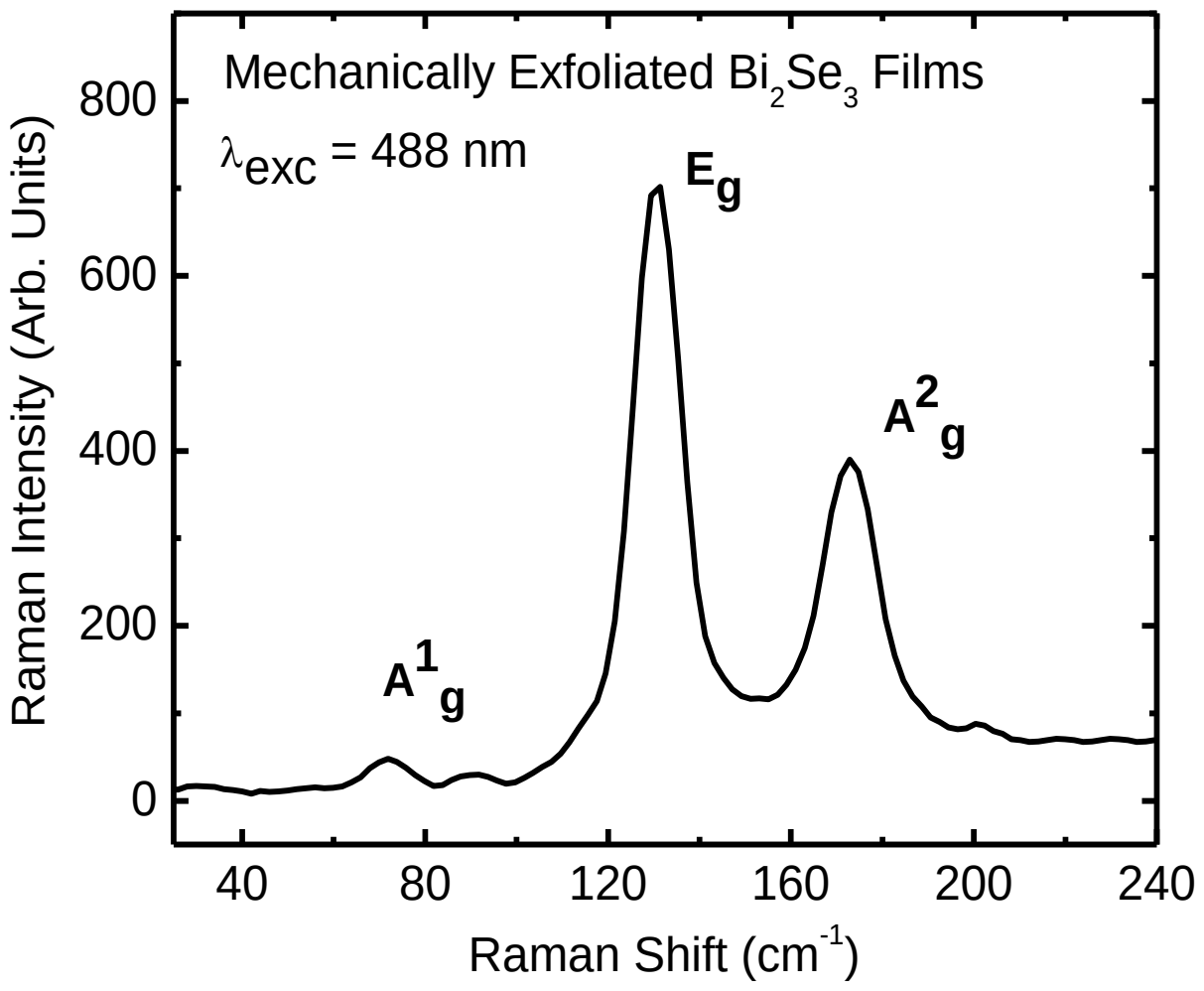

**Figure 1** Raman spectrum of $Bi_2Se_3$ few-quintuple film.

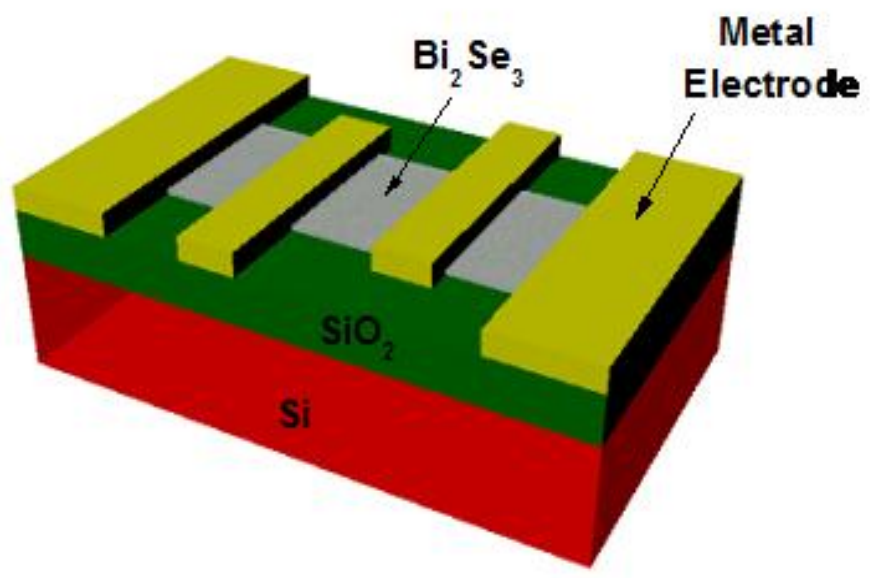

**Figure 2** Schematic of the four-terminal device with the $Bi_2Se_3$ thin-film channel.

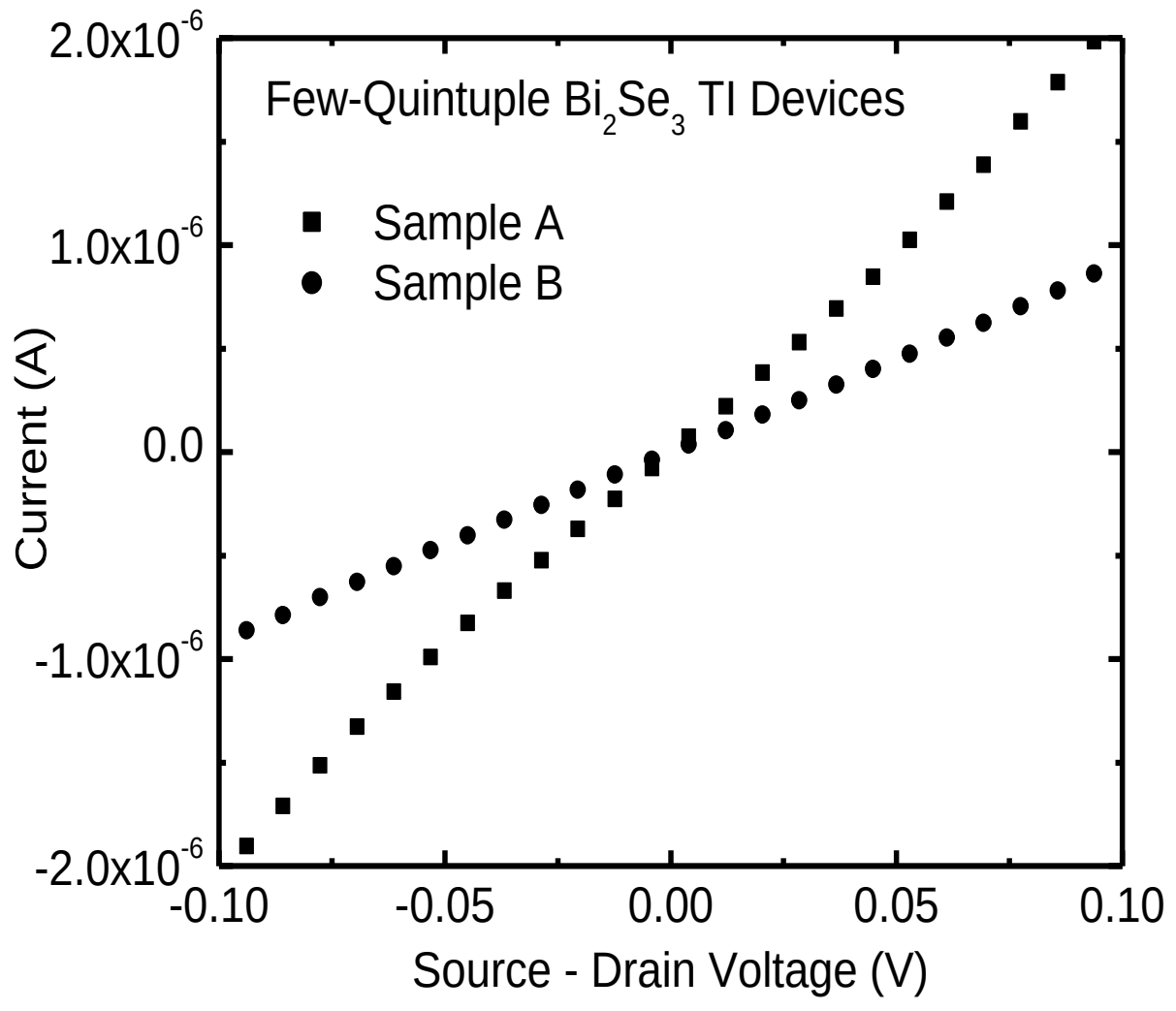

**Figure 3** Device current-voltage characteristics.

**2 Experimental Procedures** $Bi_2Se_3$ crystals can be visualized in terms of a layered structure with each layer referred to as a quintuple. Each quintuple consists of five atomic planes arranged in the sequence of -Se–Bi–Se–Bi–Se-. The coupling is strong between the atomic planes within one quintuple but the quintuples are only weakly bonded to each other by the van der Waals forces. This weak bonding allows one to obtain thin films via the "graphene-like" mechanical exfoliation. The details of the process, which we developed for a similar $Bi_2Te_3$ crystal, were reported elsewhere [6]. The exfoliation from the bulk samples allowed us to obtain crystalline films with fewer defects. The films were identified with the optical and scanning electron microscopy, and subjected to micro-Raman inspection [7].

The Raman measurements were performed at room temperature in the backscattering configuration (Renishaw inVia). The spectra were excited with a visible laser light ($\lambda$ = 488 nm) and recorded through a 50× objective. Since $Bi_2Se_3$ films have low thermal conductivity a special care was taken to avoid local heating and melting during the measurements. To improve the signal-to-noise ratio we accumulated spectra from several spots and then averaged them. Figure 1 presents Raman spectrum of $Bi_2Se_3$ film indicating its crystallinity and quality. The obtained $Bi_2Se_3$ films, with the thickness of tens of quintuples, were *n*-type with the carrier density ~$10^{18}$ $cm^{-3}$ at room temperature. The defect chemistry in $Bi_2Se_3$ is dominated by the charged Se vacancies, which act as electron donors resulting in *n*-type behaviour. The room-temperature mobility in such films is on the order of ~200 $cm^2V^{-1}s^{-1}$.

The schematic of the four-terminal device for the noise studies is shown in Figure 2. The devices were fabricated using the electron beam lithography, evaporation and lift-off process. The distance between the Ti/Au contacts ranged from 1 to 3 μm. The current – voltage characteristics were measured in the four-terminal configuration to reduce the effects of the contacts. The devices had linear characteristics in the low-bias regime as seen in Figure 3.

**3 Results and Discussion** The noise measurements were carried out with a spectrum analyzer (SR 770) at ambient conditions. Figure 4 shows typical noise spectral density for $Bi_2Se_3$ thin films. One can see that the measured excess noise is of pure 1/*f* type. The noise spectral density $S_I$ increases with increasing $V_{SD}$. Figure 5 shows the depe-ndence of 1/*f* noise spectral density measured at 1 Hz, as a function of the source – drain current. The noise density, $S_I$, follows the quadratic dependence on the current except for the knee at a very small current ($I$<5×10$^{-7}$ A). The physical origin of the knee can not be established at this stage of investigation. The noise spectral density for this device increased from ~10$^{-22}$ to 10$^{-18}$ A$^2$/Hz with the current increasing from ~10$^{-7}$ to 10$^{-5}$ A. The obtained $S_I$ values for $Bi_2Se_3$ films are important for planning electron transport experiments with topological insulators.

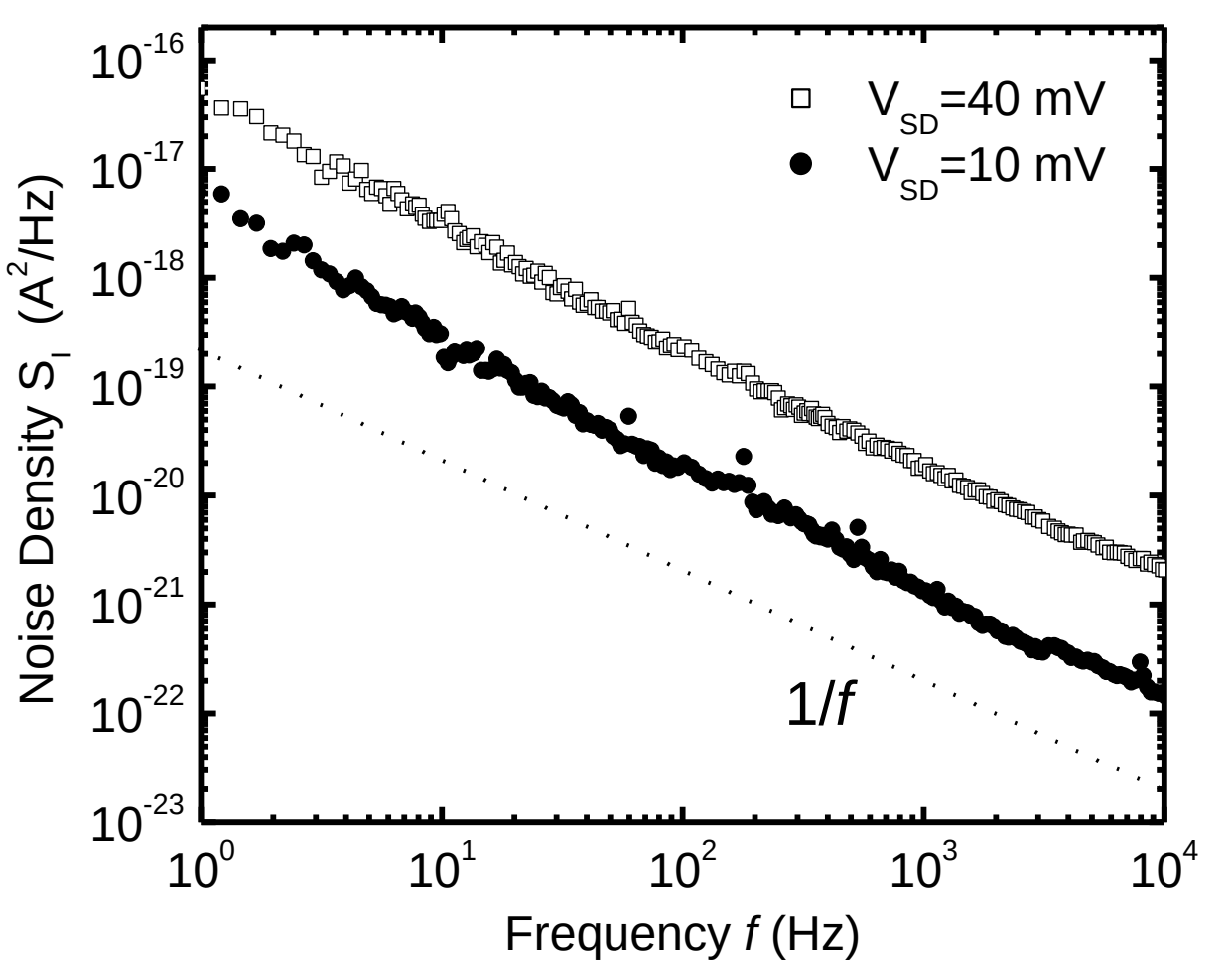

**Figure 4** Noise spectral density as a function of frequency for the conducting channel made from topological insulator. The 1/*f* spectrum is shown for comparison.

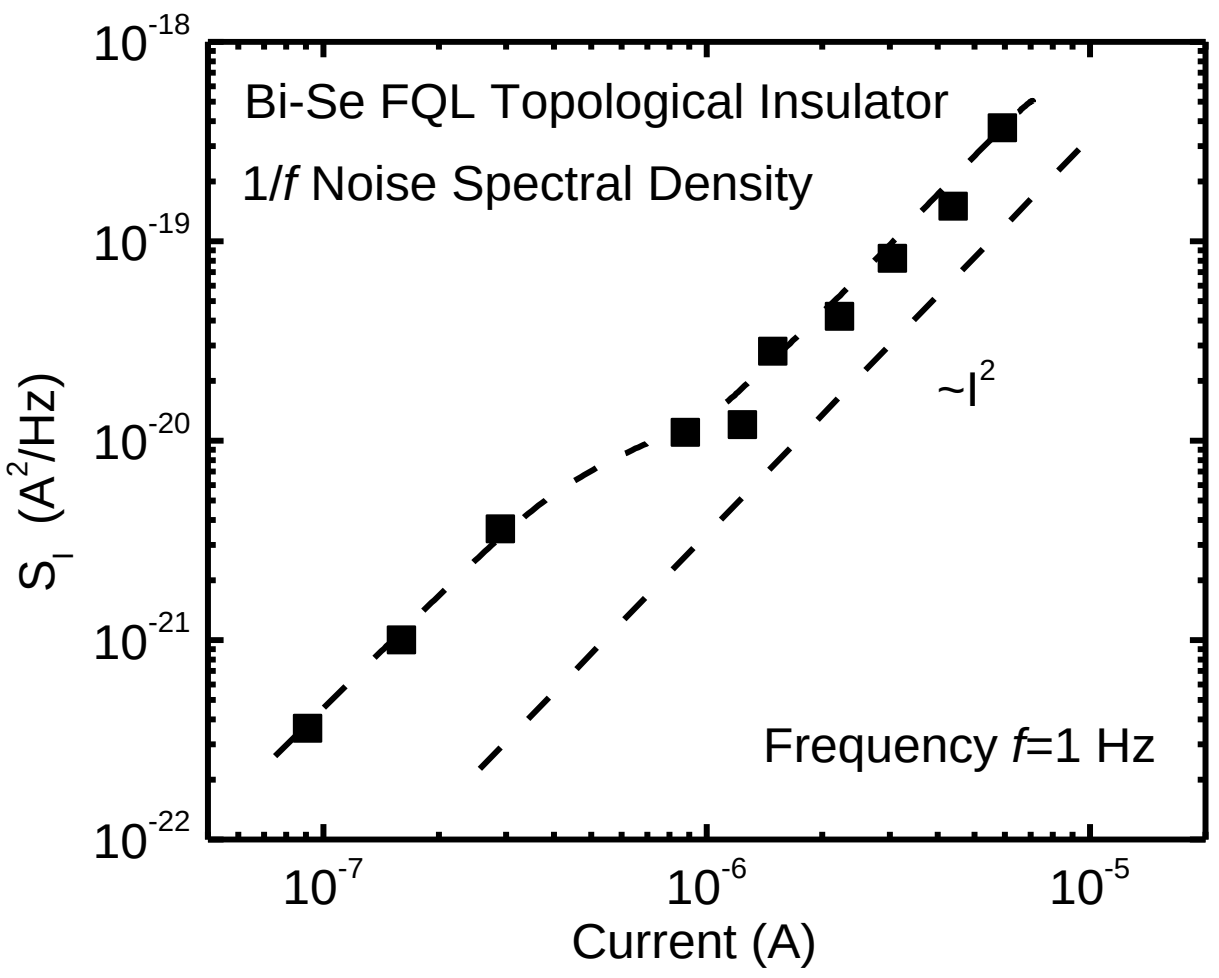

**Figure 5** Noise spectral density in the device with the channel made from $Bi_2Se_3$ topological insulator as a function of the drain-source current. The data are shown for the frequency of 1 Hz.

It has been long debated whether 1/*f* noise is produced by the mechanism distributed throughout the volume of the film (bulk effect) or localized at the sample surface (surface effect) [4-5]. The surface noise mechanism is usually linked to number-of-carriers fluctuations on the surface or in the surface oxide like in MOSFETs [8]. The bulk noise mechanism can be either due to the mobility [9] or number-of-carriers fluctuations [10]. While it is commonly accepted that in metals the noise is due to the mobility fluctuations in the bulk [5], differentiation between the number of carriers and mobility fluctuations and between the volume and surface origin of noise in semiconductors is always a problem. Recently, relatively low levels of 1/*f* noise have been reported for graphene devices [11-14]. This is somewhat unexpected for the devices where the channel consists of just one or two atomic layers of the material and, thus, represents essentially a surface ultimately exposed to the traps in the oxide layers [14].

Since the conducting surface states in topological insulators are protected against scattering by the time-reversal symmetry one can expect a possible suppression of the mobility-fluctuation noise. In this sense, topological insulators may fall into a completely different category from other materials. At the same time, the strong suppression can only be expected if the volume transport is eliminated. The latter would require further improvements in material quality and fine-tuning of the Fermi energy position with the external gates.

**4 Conclusions** We studied the low-frequency 1/*f* noise in four-terminal devices made of $Bi_2Se_3$ thin-film topological insulators. It was established that the current fluctuations in this type of materials had the pure 1/*f* spectral density for the frequency below 10 kHz. The noise spectral density followed the quadratic dependence on current. Whereas in our samples, both volume and surface mechanisms of 1/*f* noise might play a role, we suggest a possibility of suppression of 1/*f* noise in electronic devices made from topological insulators where a pure surface charge transport regime is achieved.

**Acknowledgements** The work at UCR was supported by DARPA – SRC Center on Functional Engineered Nano Architectonics (FENA). The work at RPI was supported by NSF Smart Lighting Engineering Research Center and I/UCRC "CONNECTION ONE."